\documentclass[runningheads]{svmult}

\usepackage{makeidx}   % allows index generation
\usepackage{graphicx}  % standard LaTeX graphics tool
\usepackage{subeqnar}  % subnumbers individual equations
\usepackage{multicol}  % used for the two-column index
\usepackage{physprbb}  % modified textarea for proceedings,
\makeindex             % used for the subject index
\begin{document}
\title*{Optical Diffuse Light in Nearby Compact Groups}
\toctitle{Optical Diffuse Light in Nearby Compact Groups}
% allows explicit linebreak for the table of content
%
%
\titlerunning{Diffuse light in groups}
% allows abbreviation of title, if the full title is too long
% to fit in the running head
%
\author{Cl\'audia Mendes de Oliveira\inst{1}
\and Cristiano Da Rocha\inst{1}
\and Carlos R. Raba\c ca\inst{2}
\and Daniel N. E. Pereira\inst{2}
\and Michael Bolte\inst{3}}
\authorrunning{C. Mendes de Oliveira et al.}
% if there are more than two authors,
% please abbreviate author list for running head
%
%
\institute{Instituto de Astronomia, Geof\'{\i}sica e Ci\^encias Atmosf\'ericas, U. de S\~ao Paulo, Brazil
\and Obs. do Valongo, U. Federal do Rio de Janeiro, Brazil
\and Lick Observatory, Board of Studies in Astronomy and Astrophysics, U. of
California, Santa Cruz, USA}

\maketitle              % typesets the title of the contribution

\begin{abstract}

 Analyses of B and R band observations of four compact
groups reveal the presence of a considerable amount of diffuse,
intergalactic light in two of them (HCG 79 and HCG 95).  The diffuse
component is presumably due to stellar material that has been tidally
stripped from the galaxy group members.  A new approach is used to
measure this diffuse background light, using wavelet techniques for
detecting low surface brightness signals.  The diffuse light component
has a mean colour of $(B-R)$ = 1.4 -- 1.5$\pm$0.1 and it comprises the
following fractions of the total group light in the B band: 18\%, 12\%,
3\% and 0\% for groups HCG 95, HCG 79, HCG 92 and HCG 88, respectively.
The diffuse light content of a group may represent an efficient tool
for the determination of how long groups have been together in a compact
configuration.

\end{abstract}

\section{Introduction}
  A simple visual inspection of compact groups of galaxies from
Hickson's catalogue (Hickson 1982) shows that several of them contain
diffuse background light.  This diffuse component is presumably due
to stellar material that has been tidally stripped from the galaxy
group members.  The structure of the background light is, however,
often obscured by the light from the group galaxies.  We have developed
a new method to isolate the contribution from the background light in
groups by using the {\it \`a trous} wavelet method. We apply it for
four compact group galaxies and describe the preliminary results in the
present contribution.

\section{Sample and Analysis}
  We studied a subsample of four groups from the Hickson compact
group catalogue (Hickson 1982).  The data for this analysis come from $B$
and $R$ images obtained at different telescopes and instruments. Table
1 summarizes the observations.

Details of each group studied are given in the following.

\begin{itemize} \item[$\bullet$] HCG 79 or Seyfert's sextet: This is one
of the most famous, as well as the densest, compact group of galaxies. It
is composed of four galaxies at a redshift of z=0.0145 and one foreground
galaxy at z=0.0661.

\item[$\bullet$] HCG 88: This group is composed of four spiral galaxies
lying along a line as projected on the sky, with a 2D velocity dispersion
of only 26 km s$^{-1}$. It is at z=0.0201.

\item[$\bullet$] HCG 92 or Stephan's quintet: the core of this group is
composed of one elliptical system and three spiral galaxies at a mean
redshift of z=0.0215. The fifth galaxy of the quintet is a foreground
galaxy with a velocity of $\sim$ 800 km/s.  One of the four accordant
members of the group (NGC 7318B) has a velocity that is $\sim$ 1000
km/s higher than the mean for the other three galaxies of the group.
Strong tidal filaments are observed in this group.

\item[$\bullet$] HCG 95: this quartet is the most distant group in our
sample, at a redshift of z=0.0396. One of the members of this group is an
on-going gas-rich merger with two tidal tails, which may be, in addition,
in interaction with the elliptical galaxy of the group (NGC 7609).

\end{itemize}

The images were reduced using IRAF and they were then analyzed using
a signal processing technique based on the {\it \`a trous} wavelet
transform.  Such a technique (multiscale vision model, Bijaoui and Ru\'e
1995) is well suited for analyzing complex systems composed of structures
with different sizes.  An example of various ``levels'' of the wavelet
analysis is shown in Fig. 1

\begin{figure}[]
\begin{center}
\end{center}
\caption[]{B image of HCG 79 decomposed into six ``levels'' with the
\`a trous algorithm.  Structures with different size are seen in each
level.  The panels show the detected and reconstructed objects at each
characteristic scale, from small to large, ordered from left to right
and from top to bottom.}
\label{eps1}
\end{figure}

\begin{table}
\caption{Observations of compact groups}
\begin{center}
\renewcommand{\arraystretch}{1.4}
\setlength\tabcolsep{10pt}
\begin{tabular}{llllll}
\hline\noalign{\smallskip}
Group & Tel/Instr. & Exp(B,R) & \% of B &                  $<$SB$_{B}>$ & B-R  \\
 & & seconds & \& R light & mag   \\
\noalign{\smallskip}
\hline
\noalign{\smallskip}
HCG 79 & CFHT/SIS & 2700,1800 & 13$\pm$1,12$\pm$1 & 25.5$\pm$0.1 & 1.5$\pm$0.1\\
HCG 88 & CFHT/MOS & 3600,4800 & \,~~~~~0,~~~~~0    & ~~~~~-- & ~~~~~--  \\
HCG 92 & Gemini/GMOS & \,~900,\,~~-- & \,~3$\pm$2,\,~~~-- & ~~~~~-- & ~~~~~-- \\
HCG 95 & CFHT/MOS  & 2400,3000 & 18$\pm$1,18$\pm$1 & 26.2$\pm$0.1 & 1.4$\pm$0.1\\ 
\hline
\end{tabular}
\end{center}
\label{Tab1b}
\end{table}

\section{Results}

Our preliminary results are summarized in columns 4 to 6 of Table 1.
The percentage of the total group light that is in the background halo
of diffuse light is listed in column 4, for the B and R band images.
The mean surface brightness and colour of the background light are listed
in columns 5 and 6.

The errors were obtained based on simulations of galaxies and
background light (with different shapes) using the task {\it artdata}
in IRAF. Following our simulations, we were able to determine errors
for the surface brightnesses, magnitudes, and fraction of total light
in the diffuse background component.

\section{Notes on individual groups}
 
Fig. 2 shows a B image of the group HCG 79 with the contour plots of
the diffuse light overplotted (the shape of the diffuse light component
for the R image is very similar to that measured in the B image).
The diffuse light profile of this group is irregular at all radii. The
colour of the diffuse light is similar to the colours of the outskirts of
the galaxies (or colours of low-luminosity galaxies which could possibly
have been stripped).  It is interesting to note that the HI contours
(Williams et al. 1991) of HCG 79 show similar shape to the optical
diffuse light envelope.

\begin{figure}[]
\begin{center}
\end{center}
\caption[]{
B image of H79 with superimposed diffuse light intensity curves
(continuous lines). The brightest contours are, from the center outwards,
25.0, 25.25, 25.5, 25.9, 26.4 and 27.5 mag/arcsec$^2$.  The  center of
the diffuse light component and the geometric center of the group are
shown with ``$\times$'' and ``+'', respectively.
}
\label{eps2}
\end{figure}

For HCG 95, the optical diffuse light we detected represented 18\% of
the total light of the group and for HCG 88, we could detect no optical
diffuse light.

The analysis of an R image of the Stephan quintet, obtained with Gemini
in July 2002 shows that although there are several tails and other sharp
low-luminosity features, there does not seem to exist an envelope of light
around the group. Instead, the diffuse background light is concentrated
in a position north of the pair NGC 7318B/D, in between the two northern
tails which emanate from these galaxies.

\section{Summary, Conclusion and Perspectives}
 
Previous claims for a lack of diffuse optical light in compact groups
(Rose 1979; Pildis, Bregman, \& Schombert 1995) were based on data
insensitive to low surface brightness features or not fully processed
to reveal them.  Our analysis of the $B$ and $R$ images of four compact
groups using wavelet analysis indicated the presence of background light
in three of them.

 Important information about the influence of interaction processes
in galaxy evolution can be provided by the detection of diffuse light.
Compact groups have a high surface density compared to the field
and most of the $N>4$ groups are probably bound and virialized. The
presence of light that cannot be explained by the overlapping envelopes
of individual galaxies indicates that tidal encounters have already
stripped the galaxies of significant mass. The stripped material should be
distributed in the group potential in the form of a diffuse intergalactic
medium of stars and clusters.  The presence of significant halos around
dense systems constitute proof that the galaxies have been interacting,
that they have had significant dynamical evolution and provides a test
of group models.

  In the particular case of the two groups with the largest amount
of diffuse light studied here, the presence of a diffuse halo indicates
that  the member galaxies have been together for at least a few crossing
times. On the other hand, the irregular shapes of the diffuse light
envelopes suggest that the members have been together for a relatively
short timescale.

We plan to combine the measures of the fraction of the total light in
diffuse halos and the shapes of the halos with n-body predictions in
order to obtain estimates for the dynamical ages of dense groups.

\section{Acknowledgments}
C.M.dO. and C.D.R. would like to acknowledge funding from FAPESP (grant
No. 96/08986-5). They also thank PRONEX, the Alexander von Humboldt
Foundation, ESO and the conference organizers for making possible the
attendance to the conference. CRR acknowledges funding from FAPERJ
(E-26/171.085-99). We are grateful to Laerte Sodr\'e Jr. and Eduardo
Cypriano for the joint Gemini proposal of HCG 92.

%INDEX%%%%%%%%%%%%%%%%%%%%%%%%%%%%%%%%%%%%%%%%%%%%%%%%%%%%%%%%%%%%%%%
% Please check with the editor of your book whether he plans to
% include a "mutual" subject index - if so, please code your entries
% in the standard syntax. For your own purposes you may print your
% "personal" index by using the following commands:
%
%\clearpage
%\addcontentsline{toc}{section}{Index}
%\flushbottom
%\printindex
%%%%%%%%%%%%%%%%%%%%%%%%%%%%%%%%%%%%%%%%%%%%%%%%%%%%%%%%%%%%%%%%%%%%%


\begin{thebibliography}{8.}
\addcontentsline{toc}{section}{References}

\bibitem{bij95} A. Bijaoui, F. Ru\'e: Signal Processing \textbf{46}, 345 (1995)
W. Frank, A. Seeger: Appl. Phys. A \textbf{3}, 66 (1988)

\bibitem{hic82} P. Hickson: ApJ \textbf{255}, 382 (1982)

\bibitem{pil95a} R. A. Pildis, J.N. Bregman, A.E. Evrard: ApJ \textbf{443}, 514 (1995)

\bibitem{pil95b} R.A. Pildis, J.N. Bregman, J.M. Schombert: AJ \textbf{110}, 
1498 (1995)

\bibitem{ros79} J.A. Rose: ApJ \textbf{231}, 10 (1979)

\bibitem{wil91} B.A. Williams, P.M. McMahon, J.H. van Gorkom: AJ \textbf{101}, 1957 (1991)

\end{thebibliography}
\end{document}